\documentclass{caosp306}

\usepackage{graphicx}
\usepackage[dvipsnames]{xcolor}
\usepackage{natbib}
\articleNo{123}
\pubyear{2020}
\volume{50}
\volnumber{1}
\firstpage{350}
\received{August 12, 2019}
\accepted{September 2, 2019}

\def\BibTeX{{\rm B\kern-.05em{\sc i\kern-.025em b}\kern-.08em
             T\kern-.1667em\lower.7ex\hbox{E}\kern-.125emX}}

\begin{document}

%
\hauthor{I.\,Jankov, D.\,Ilic}

\title{Narrow lines correlations in an SDSS sample of
type 1 quasars}


%
\author{
        I.\,Jankov
      \and 
        D.\,Ilic
       }

%
\institute{
           Department of Astronomy, Faculty of Mathematics, University of Belgrade, Studentski trg 16, 11000 Belgrade, Serbia 
          }

\date{July 31, 2019}

\maketitle

\begin{abstract}
Investigation of quasar emission line properties and relationships between spectral parameters is important for understanding the physical mechanisms that originate inside different regions of the active galactic nuclei. In this paper, we investigate the optical spectral parameters of type 1 quasars taken from  the Sloan Digital Sky Survey Data Release 7 Quasar Catalog \citep{2011ApJS..194...45S}. Spectral parameters, such are equivalent widths and full widths at half maximum of both narrow and broad lines are taken into account. We perform the analysis of correlation matrix and principal component analysis of our sample. We obtain that the narrow line Baldwin effect is significant enough and deserves further investigation. We provide the correlation coefficients and slope values for Baldwin effect in several narrow lines.
\end{abstract}

%
\

\section{Introduction}
\label{intr}
Active galactic nuclei (AGN) or quasars are extremely bright objects with rapidly accreting supermassive black hole at their center. They exhibit a wide range of spectral characteristics, which can give us valuable information about the physical conditions in these extreme environments. Large surveys of quasars showed that there are some correlations between their spectral parameters. Most notable correlations are anti-correlation between [O\,III] and optical Fe\,II line strength, as well as the anti-correlation between full width at half maximum (FWHM) of broad H$\beta$ line and ratio of optical Fe\,II line and broad H$\beta$ line equivalent widths ($\mathrm{R_{\mathrm{Fe\,II}}}$). These correlations were investigated using principal component analysis (PCA) and represented in the parameter space known as Eigenvector 1 \citep[see e.g.][]{2000ApJ...536L...5S}. In addition to the Eigenvector 1 correlations, the strength of some spectral lines appears to decrease with luminosity of the underlying continuum. This trend was first detected in C IV line and is known  as the Baldwin effect \citep{1977ApJ...214..679B}.


There is distinction between the Baldwin effect for narrow and broad emission lines because these lines are associated with different regions of AGN with different geometrical and dynamical properties. In case of broad emission lines, it has been studied in great detail \citep[see e.g][]{2002ApJ...581..912D, 2007ASPC..373..355S, 2012MNRAS.427.2881B}, but its  physical origin is still a matter of debate  \citep[see e.g.][]{2017A&A...603A..49R}. One possible explanation for Baldwin effect that is associated with broad emission lines, is that the continuum shape is luminosity-dependent, in a way that more luminous objects have softer UV/X-ray spectra and this results in reduced ionization and photoelectric heating in the broad line region (BLR) gas \citep[e.g.][] {1992ApJS...80..109B,1998ApJ...507...24K}. The narrow line Baldwin effect is still not entirely understood \citep{2007ASPC..373..355S}, and has been studied by several authors in the past. For example, \cite{2010ApJS..189...15K} did a careful spectral measurement of a sample of $\sim$300 AGN and noticed a strong anti-correlation between the narrow H$\beta$ and [O III] lines with continuum luminosity, with the correlation coefficients of  -0.36 and -0.43, respectively. The explanation of the effect could be due to different scale of narrow line region (NLR) compared to the BLR, or due to the extinction of the continuum luminosity by dust located between  the BLR and NLR \citep[e.g.][]{2016MNRAS.461.4227H}, or due to  some other factors that also need to be taken into account \citep[for more details see][]{1992ApJS...80..109B, 1999ApJ...514...40M, 2002MNRAS.337..275C,2002ApJ...581..912D, 2004ApJ...614..558N, 2006A&A...453..525N, 2013ApJ...762...51Z, 2007ASPC..373..355S}. 

The Sloan Digital Sky Survey (SDSS) quasars were previously  investigated for the Baldwin effect in narrow lines by  \cite{2013ApJ...762...51Z}, who studied a large sample of broad-line (i.e. type 1), radio-quiet AGN taken from SDSS Data Release 4. They found that  narrow lines show a similar Baldwin effect with slope value of approximately -0.2, which could be explained with the combination of continuum variation and a lognormal distribution of the luminosity. In addition they show that there is no evidence for a relationship between Baldwin effect slope and ionization potential of the narrow lines \citep{2013ApJ...762...51Z}. However, in their analysis they did not consider the host galaxy starlight contamination, which overestimates continuum luminosity, especially for low luminosity quasars \citep[e.g.][]{2006ApJ...644..133B, 2011ApJS..194...45S}, which is resulting in steeper Baldwin effect slope.

In this paper, we report the results of our analysis based on the correlation matrices and PCA on a sample of type 1 quasars taken from the SDSS Data Release 7 (SDSS DR7) quasar catalog \citep{2011ApJS..194...45S}, with a special focus on the narrow line Baldwin effect, which was not investigated before in case of this catalog.

\

\section{Data Reduction and Analysis}

 We use the measured spectral line properties of the SDSS quasars from \cite{2011ApJS..194...45S} catalog, and we select our sample putting the following constraints: the signal to noise ratio S/N\,\textgreater\,10 and redshift z\,\textless\,0.39. The higher S/N ratio was set in order to have more precisely measured quantities \citep{2011ApJS..194...45S}  and redshift was limited so both H$\alpha$ and H$\beta$ spectral ranges are covered. This has resulted with a sample of 2,\,224 type 1 quasars. All objects in the sample have the following spectral parameters measured:  the FWHM of broad H$\alpha$ and H$\beta$ lines; equivalent widths (EW) of narrow H$\alpha$, H$\beta$, [O\,III]\,$\lambda$5007\,\angstrom, [S\,II]\,$\lambda\lambda$6718,\,6732\,\angstrom\, and [N\,II]\,$\lambda$6585\,\angstrom\, lines; continuum luminosity at 5100\,\angstrom\, ($L_{5100}$), $R_{\rm FeII}$ and luminosity ratio of [O\,III] and narrow H$\beta$ line. We emphasize that the final number of objects is obtained so that there are no null values for all spectral parameters.


\begin{figure}
\vspace{-0.5cm}
\centerline{\includegraphics[width=13cm]{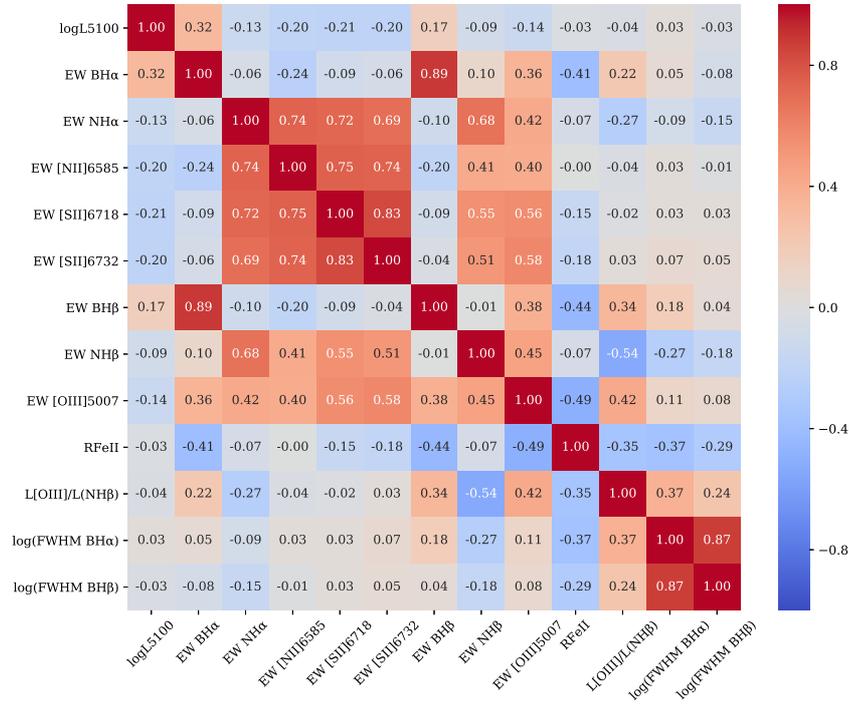}}
\vspace{0.4cm}
\caption{Correlation matrix of the entire sample. The color bar on the right represents the color associated with a correlation coefficient value. Red is associated with correlations and blue is associated with anti-correlations. The numbers represent Spearman correlation coefficients.}
\label{corr}
\end{figure}

To analyze the correlations between quasar optical spectral parameters, we constructed a correlation matrix of our sample using Spearman correlation coefficients. The correlation matrix served as a useful tool for the selection of parameters later for PCA, but also for the analysis of correlation coefficients separately from the PCA results. 

PCA is sometimes considered to be an unsupervised machine learning algorithm, because it can be used on unlabeled datasets, i.e. it can be applied even if we don't have a target variable that we want to predict \citep[for more details on PCA see e.g.][]{1992ApJS...80..109B, 1999ASPC..162..363F, 2004AJ....127.1799G}. 

In this work, the PCA was applied using the scikit-learn Python library where dimensionality reduction is conducted using Singular Value Decomposition of the data to project it to a lower dimensional space. In this way, only a few components are needed to completely describe total variance in our dataset and each component is representation of the dominant trends in the data.

To calculate the Baldwin effect slope, we performed linear regression for each narrow emission line in our sample. For further analysis of the Baldwin effect, the continuum luminosity was divided onto 9 bins of length 0.2 dex, starting from log\,$L_{5100} = 43.75$ and ending with log\,$L_{5100} = 45.55$. The last bin was excluded from the analysis because it contained only one object and it was not representative enough to be taken into account. Finally, to test the effect of the starlight contamination on the Baldwin effect, we corrected the continuum luminosity using the empirical fitting formula of the average host contamination given in \cite{2011ApJS..194...45S}, see their Eq. 1, and obtained the pure continuum luminosity of the quasar. The luminosity at 5100 \AA \, has been corrected up to log\,$L_{5100} < 45.053$, since above this value no correction is needed.

\

\section{Results and Discussion}

Fig. \ref{corr} gives the correlation matrix of all spectral parameters in our  sample of type 1 quasars. The most evident result is the
presence of correlations between parameters which populate Eigenvector 1 (E1) parameter space, which was well established and investigated previously \citep[for a review see][]{2000ARA&A..38..521S}: the anti-correlations between the [O III] equivalent width and the Fe II strength $R_{\rm Fe II}$ ($r=-0.49$), and the FWHM of broad H$\beta$ line and $R_{\rm Fe II}$ ($r=-0.29$). The strongest correlations are found between the equivalent widths of broad H$\alpha$ and H$\beta$ lines ($r=0.89$) as well as between their narrow components ($r=0.68$), which is an expected result.

\begin{figure}
\centerline{\includegraphics[width=13cm,clip=]{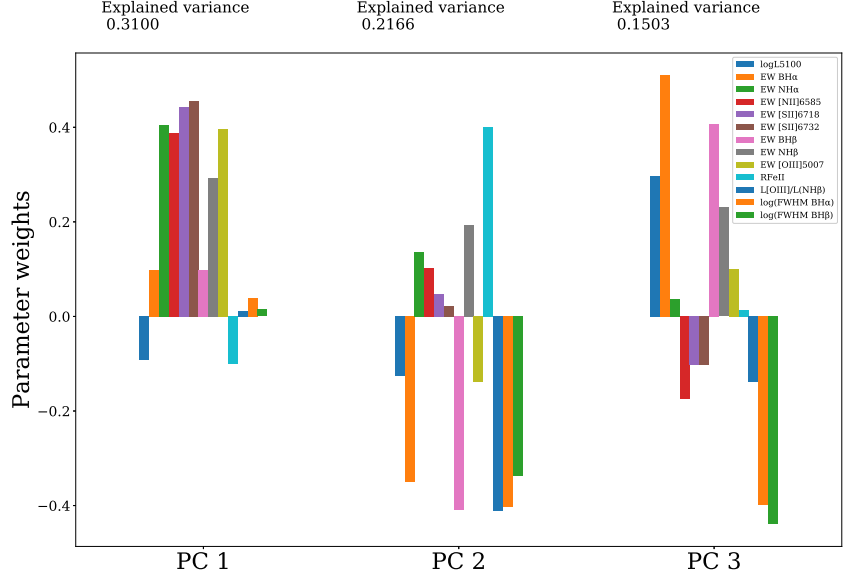}}
\caption{Results of the PCA, applied on the whole sample of 2,224 quasars, presented with bar charts. The three most significant principal components (eigenvectors) are labeled on the horizontal axis and on the vertical axis they are described in terms of the parameter weights. Each principal component is a linear combination of our 13 parameters, where the weights represent the components of the eigenvector. Note here that if the weight of at least two parameters is of the same sign, they are correlated and if the weights are of the opposite sign, they are anti-correlated. The parameters with the largest absolute value of the weights show the largest correlations/anti-correlations. The extent to which each component is describing the data sample is presented with the fraction of the total variance in the data sample and this number is indicated on top of each components bar chart. The width of the bars is arbitrary and serves only to illustrate the results.}
\label{pca}
\end{figure}

However, the matrix clearly reveals the apparent anti-correlation between continuum luminosity at 5100 \angstrom\, and equivalent widths of several narrow lines, which include: [O\,III]\,$\lambda$5007\,\angstrom, [N\,II]\,$\lambda$6585\,\angstrom, [S\,II]\,$\lambda\lambda$6718,\,6732\,\angstrom, narrow H$\alpha$ and narrow H$\beta$. This anti-correlation represents the Baldwin effect for narrow lines in AGN. While Baldwin effect for broad lines in AGN is well established, narrow line Baldwin effect is still under investigation and it may result from different physical mechanisms. 

The results of the PCA, which are presented in Fig. \ref{pca}, are also indicating the presence of these anti-correlations. Furthermore, PCA suggests that this is the dominant trend in our data sample because the relationship between continuum luminosity and narrow line equivalent widths is represented with the first principal component (PC\,1). PC\,1 explains 31\% of the variance in our data sample. The second principal component (PC\,2) is dominated by the anti-correlation between $R_{\rm FeII}$ and FWHM of broad H$\beta$ \citep{2000ApJ...536L...5S} and anti-correlation between $R_{\rm FeII}$ and [O\,III]\,$\lambda$5007 \citep[discovered by][]{1992ApJS...80..109B}. One other correlation that is very prominent on PC2 is the correlation between the ratio of [O\,III] and H$\beta$ luminosity L([O\,III])/L(H$\beta$) and FWHM of broad H$\alpha$ which was recently discussed by \cite{2019MNRAS.487.3404B}. 

\begin{figure}
\centerline{\includegraphics[width=13cm,clip=]{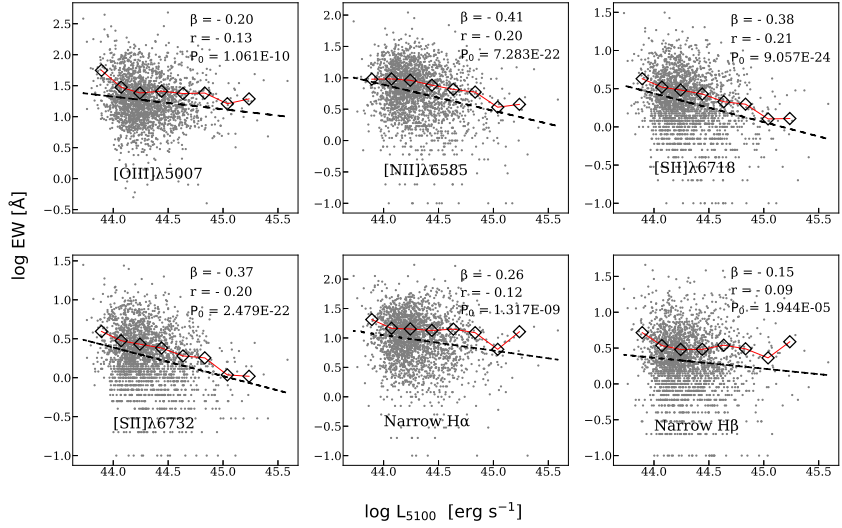}}
\caption{The Baldwin effect (log\,$\mathrm{L_{5100}}$ - log\,EW) in case of six narrow lines for all data in our sample (gray dots). Slopes ($\beta$) are obtained from linear regression (black dashed line) applied on the whole sample. Diamond markers (connected with red full line) represent points that resulted from continuum luminosity binning. The correlation coefficient (r) and the p-value of the null hypothesis ($P_{0}$) is also given on each plot.}
\label{Beff}
\end{figure}

The summary of correlation coefficients with p-values of the null hypothesis for all narrow lines, along with the values of slopes obtained from linear regression is given in Table \ref{t1}. These values are also presented on Fig. \ref{Beff}.  Table \ref{t1} also lists the ionization energy and critical density in case of  forbidden lines. The strongest Baldwin effect appears to be associated with [N\,II] and [S\,II] lines, however the p-values suggests that these anti-correlations are significant in all narrow lines. The  Baldwin effect of the H$\alpha$ and H$\beta$ lines shows the lowest correlation with the smallest significance.

We find that the slope for [O\,III] and H$\alpha$ are in agreement with the findings of \cite{2013ApJ...762...51Z}, however, for 
[N\,II] and [S\,II] lines the slope is is much steeper ($\beta=-0.41$ and $\beta=-0.38$, respectively) than previously reported  by \cite{2013ApJ...762...51Z} , who received the value of $\beta=-0.10$ and $\beta=-0.20$.  This could be the result of the selection effect, since our sample contains low-redshift objects, and both radio loud and radio quiet objects. 



\begin{table}[t]
\small
\begin{center}
\caption{($\beta$) - the log EW - log\,$L_{5100}$ slope; ($r$) - correlation coefficient; ($P_{0}$) - p-value of the null hypothesis; ($\chi_{ion}$) - line ionization energy in eV; ($n_{c}$) - logarithm of the line critical density in cm$^{-3}$.}
\label{t1}
\begin{tabular}{lccccc}
\hline\hline
{Emission line}  &{$\beta$}& {$\it r$} & {$\it P_{0}$} &{$\it \chi_{ion}$} & {log\,$n_{c}$}\\
\hline
$\mathrm{[O\,III]\,\lambda 5007}$ & -0.20$\pm$0.03  & -0.14  & $1.06\mathrm{E}{-10}$& 35.11 & 5.80 \\
$\mathrm{[N\,II]\,\lambda 6585}$  & -0.42$\pm$0.04  & -0.20  & $7.28\mathrm{E}{-22}$ &14.50&4.82\\
$\mathrm{[S\,II]\,\lambda 6718}$  & -0.38$\pm$0.03  & -0.21  & $9.06\mathrm{E}{-24}$ &10.36&2.30\\
$\mathrm{[S\,II]\,\lambda 6732}$  & -0.38$\pm$0.03  & -0.20  & $2.48\mathrm{E}{-22}$&10.36&2.30 \\
Narrow H$\alpha$                           & -0.26$\pm$0.04  & -0.13  & $1.32\mathrm{E}{-09}$&13.60&infinity\\
Narrow H$\beta$                             & -0.15$\pm$0.04  & -0.09  & $1.94\mathrm{E}{-05}$&13.60&infinity\\
\hline\hline
\end{tabular}
\end{center}
\end{table}

\begin{figure}
\centerline{\includegraphics[width=11cm,clip=]{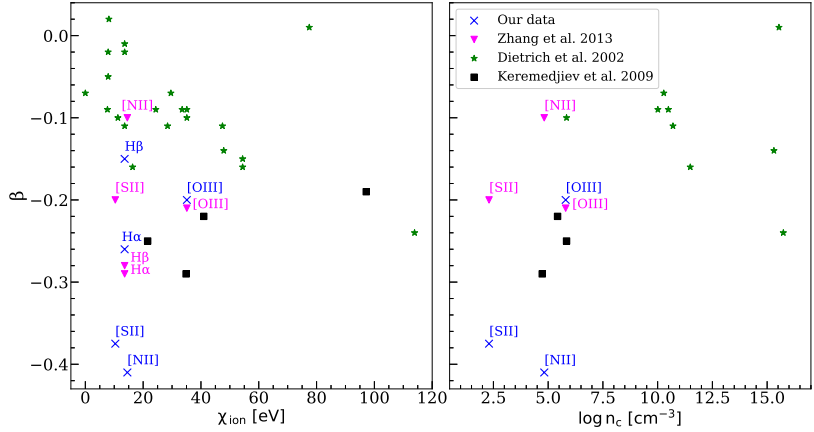}}
\caption{Dependence of Baldwin effect slope on ionization potential ($\chi_{ion}$) and critical density ($n_{c}$) of lines in question. The blue crosses represent our data and data from \cite{2013ApJ...762...51Z, 2002ApJ...581..912D, 2009ApJ...690.1105K} is represented with pink triangles, green stars and black squares, respectively.}
\label{ion}
\end{figure}

In order to quickly test if the Baldwin effect slope for narrow lines depends on the ionization potential or critical density of the lines, which was noticed to be relevant in case of the broad line Baldwin effect \citep[e.g.,][]{1999ASPC..162..351E, 2002ApJ...581..912D}, we plotted those dependencies in Fig. \ref{ion}, and compared our results with the data obtained from previous studies of narrow line Baldwin effect \citep{2013ApJ...762...51Z, 2009ApJ...690.1105K, 2002ApJ...581..912D}. The p-values of the null hypothesis of correlations in question  indicate that there is no significant correlation between Baldwin effect slopes and ionization potential ($P_{0} = 0.87$), and no correlation between Baldwin effect slopes and line critical densities ($P_{0} = 0.17$) for narrow lines. This is in agreement with findings of \cite{2013ApJ...762...51Z}. 

Furthermore, we selected from our sample a sub-sample of 405 radio-loud quasars based on detection by the FIRST survey (both lobe- and core-dominated objects are included), and a sub-sample of 1,694 radio-quiet quasars, for which the FIRST survey reported no radio emission. Preliminary results of this analysis indicate that narrow line Baldwin effect slopes are steeper for radio-loud than for radio-quiet quasars. In the case of radio-loud quasars [O\,III]\,$\lambda$5007 line has a slope ($\beta$) of -0.36, correlation coefficient ($r$) is -0.24 and p-value of the null hypothesis ($P_{0}$) is 7.48E-07. On the other hand, in the case of radio-quiet quasars the same line has $\beta$ = -0.20, $r$ = -0.14 and $P_{0}$ = 2.06E-08. For the [N\,II]\,$\lambda$6583 line the results also differ substantially, such that we have for radio-loud sample $\beta$ = -0.56, $r$ = -0.35 and $P_{0}$ = 4.85E-13 and for radio-quiet $\beta$ = -0.41, $r$ = -0.19 and $P_{0}$ = 3.02E-15.
The difference that appears in these two sub-samples may originate from the fact that in radio-loud quasars the narrow line emission may be enhanced by the jet-ISM interaction \citep{2008A&A...488L..59L} and also the continuum emission can be enhanced as the result of relativistic beaming. Further investigations are needed to confirm if this is the case.

Finally, we tested how would starlight contamination of the continuum luminosity affect the Baldwin effect in narrow lines. Using the continuum luminosity corrected for the host contribution, we still detect the Baldwin effect in all considered narrow lines, but with consistently smaller slope. The slope of the [O\,III]\,$\lambda$5007 is  decreased from -0.20 to -0.16, for  [N\,II]\,$\lambda$6585 to -0.32, for 
[S\,II]\,$\lambda$6718 and [S\,II]\,$\lambda$6732 to -0.29, for H$\alpha$ to -0.20, and for H$\beta$ to -0.12. 
The correlation coefficients remain the same.

\

\section{Conclusions}

In this paper, we reported the results of the analysis performed on the sample of 2,224 low-redshift (z\,\textless\,0.39) quasars from SDSS quasar catalog by \cite{2011ApJS..194...45S}, for which we constructed the correlation matrix and applied the PCA on their optical spectral parameters. From our results, we came to the following conclusions: (i) we confirmed the Eigenvector 1 parameter space correlations for our data sample;
ii) the Baldwin effect is present in all studied narrow lines, but with different slopes; iii) the slopes and correlation coefficients indicate that the Baldwin effect is the strongest for $\mathrm{[N\,II]\,\lambda 6585}$  and $\mathrm{[S\,II]\,\lambda\lambda 6718,6732}$ lines, whereas in the case of the H$\alpha$ and H$\beta$ lines it shows the lowest correlation with the smallest significance; iv) ionization energies needed to create the ions and critical densities of the narrow lines have no correlation with Baldwin effect slopes for narrow lines; and v) the host galaxy contamination of the continuum luminosity is steepening the Baldwin effect, but even with the corrected luminosity, the Baldwin effect remains in all narrow lines. Future investigation should be done to test if the continuum variation could be an explanation of the Baldwin effect as suggested by \cite{2013ApJ...762...51Z}.

\acknowledgements
This work is supported by the project (176001) Astrophysical Spectroscopy
of Extragalactic Objects of the Ministry of Education, Science and Technological Development of the Republic of Serbia.

\bibliography{Jankov_Ilic}

\begin{thebibliography}{24}
\expandafter\ifx\csname natexlab\endcsname\relax\def\natexlab#1{#1}\fi

\bibitem[{{Baldwin}(1977)}]{1977ApJ...214..679B}
{Baldwin}, J.~A., {Luminosity Indicators in the Spectra of Quasi-Stellar
  Objects}. 1977, {\it \apj}, {\bf 214}, 679, DOI: 10.1086/155294

\bibitem[{{Baron} \& {M{\'e}nard}(2019)}]{2019MNRAS.487.3404B}
{Baron}, D. \& {M{\'e}nard}, B., {Black hole mass estimation for active
  galactic nuclei from a new angle}. 2019, {\it \mnras}, {\bf 487}, 3404, DOI:
  10.1093/mnras/stz1546

\bibitem[{{Bentz} {et~al.}(2006){Bentz}, {Peterson}, {Pogge}, {Vestergaard}, \&
  {Onken}}]{2006ApJ...644..133B}
{Bentz}, M.~C., {Peterson}, B.~M., {Pogge}, R.~W., {Vestergaard}, M., \&
  {Onken}, C.~A., {The Radius-Luminosity Relationship for Active Galactic
  Nuclei: The Effect of Host-Galaxy Starlight on Luminosity Measurements}.
  2006, {\it \apj}, {\bf 644}, 133, DOI: 10.1086/503537

\bibitem[{{Bian} {et~al.}(2012){Bian}, {Fang}, {Huang}, \&
  {Wang}}]{2012MNRAS.427.2881B}
{Bian}, W.-H., {Fang}, L.-L., {Huang}, K.-L., \& {Wang}, J.-M., {The C IV
  Baldwin effect in quasi-stellar objects from Seventh Data Release of the
  Sloan Digital Sky Survey}. 2012, {\it \mnras}, {\bf 427}, 2881, DOI:
  10.1111/j.1365-2966.2012.22123.x

\bibitem[{{Boroson} \& {Green}(1992)}]{1992ApJS...80..109B}
{Boroson}, T.~A. \& {Green}, R.~F., {The Emission-Line Properties of
  Low-Redshift Quasi-stellar Objects}. 1992, {\it \apjs}, {\bf 80}, 109, DOI:
  10.1086/191661

\bibitem[{{Croom} {et~al.}(2002){Croom}, {Rhook}, {Corbett}, {Boyle}, {Netzer},
  {Loaring}, {Miller}, {Outram}, {Shanks}, \& {Smith}}]{2002MNRAS.337..275C}
{Croom}, S.~M., {Rhook}, K., {Corbett}, E.~A., {et~al.}, {The correlation of
  line strength with luminosity and redshift from composite quasi-stellar
  object spectra}. 2002, {\it \mnras}, {\bf 337}, 275, DOI:
  10.1046/j.1365-8711.2002.05910.x

\bibitem[{{Dietrich} {et~al.}(2002){Dietrich}, {Hamann}, {Shields},
  {Constantin}, {Vestergaard}, {Chaffee}, {Foltz}, \&
  {Junkkarinen}}]{2002ApJ...581..912D}
{Dietrich}, M., {Hamann}, F., {Shields}, J.~C., {et~al.}, {Continuum and
  Emission-Line Strength Relations for a Large Active Galactic Nuclei Sample}.
  2002, {\it \apj}, {\bf 581}, 912, DOI: 10.1086/344410

\bibitem[{{Espey} \& {Andreadis}(1999)}]{1999ASPC..162..351E}
{Espey}, B. \& {Andreadis}, S., {Observational Evidence for an
  Ionization-Dependent Baldwin Effect}. 1999, in Astronomical Society of the
  Pacific Conference Series, Vol. {\bf  162}, {\it Quasars and Cosmology}, ed.
  G.~{Ferland} \& J.~{Baldwin}, 351

\bibitem[{{Francis} \& {Wills}(1999)}]{1999ASPC..162..363F}
{Francis}, P.~J. \& {Wills}, B.~J., {Introduction to Principal Components
  Analysis}. 1999, in Astronomical Society of the Pacific Conference Series,
  Vol. {\bf  162}, {\it Quasars and Cosmology}, ed. G.~{Ferland} \&
  J.~{Baldwin}, 363

\bibitem[{{Grupe}(2004)}]{2004AJ....127.1799G}
{Grupe}, D., {A Complete Sample of Soft X-Ray-selected AGNs. II. Statistical
  Analysis}. 2004, {\it \aj}, {\bf 127}, 1799, DOI: 10.1086/382516

\bibitem[{{Heard} \& {Gaskell}(2016)}]{2016MNRAS.461.4227H}
{Heard}, C. Z.~P. \& {Gaskell}, C.~M., {The location of the dust causing
  internal reddening of active galactic nuclei}. 2016, {\it \mnras}, {\bf 461},
  4227, DOI: 10.1093/mnras/stw1616

\bibitem[{{Keremedjiev} {et~al.}(2009){Keremedjiev}, {Hao}, \&
  {Charmandaris}}]{2009ApJ...690.1105K}
{Keremedjiev}, M., {Hao}, L., \& {Charmandaris}, V., {The Mid-Infrared
  Narrow-Line Baldwin Effect Revealed by Spitzer}. 2009, {\it \apj}, {\bf 690},
  1105, DOI: 10.1088/0004-637X/690/2/1105

\bibitem[{{Korista} {et~al.}(1998){Korista}, {Baldwin}, \&
  {Ferland}}]{1998ApJ...507...24K}
{Korista}, K., {Baldwin}, J., \& {Ferland}, G., {Quasars as Cosmological
  Probes: The Ionizing Continuum, Gas Metallicity, and the
  W$_{{\ensuremath{\lambda}}}$-L Relation}. 1998, {\it \apj}, {\bf 507}, 24,
  DOI: 10.1086/306321

\bibitem[{{Kova{\v{c}}evi{\'c}} {et~al.}(2010){Kova{\v{c}}evi{\'c}},
  {Popovi{\'c}}, \& {Dimitrijevi{\'c}}}]{2010ApJS..189...15K}
{Kova{\v{c}}evi{\'c}}, J., {Popovi{\'c}}, L.~{\v{C}}., \& {Dimitrijevi{\'c}},
  M.~S., {Analysis of Optical Fe II Emission in a Sample of Active Galactic
  Nucleus Spectra}. 2010, {\it \apjs}, {\bf 189}, 15, DOI:
  10.1088/0067-0049/189/1/15

\bibitem[{{Labiano}(2008)}]{2008A&A...488L..59L}
{Labiano}, A., {Tracing jet-ISM interaction in young AGN: correlations between
  [O III] {\ensuremath{\lambda}} 5007 {\r{A}} and 5-GHz emission}. 2008, {\it
  \aap}, {\bf 488}, L59, DOI: 10.1051/0004-6361:200810399

\bibitem[{{McIntosh} {et~al.}(1999){McIntosh}, {Rieke}, {Rix}, {Foltz}, \&
  {Weymann}}]{1999ApJ...514...40M}
{McIntosh}, D.~H., {Rieke}, M.~J., {Rix}, H.~W., {Foltz}, C.~B., \& {Weymann},
  R.~J., {A Statistical Study of Rest-Frame Optical Emission Properties in
  Luminous Quasars at 2.0\&lt;=z\&lt;=2.5}. 1999, {\it \apj}, {\bf 514}, 40,
  DOI: 10.1086/306936

\bibitem[{{Netzer} {et~al.}(2006){Netzer}, {Mainieri}, {Rosati}, \&
  {Trakhtenbrot}}]{2006A&A...453..525N}
{Netzer}, H., {Mainieri}, V., {Rosati}, P., \& {Trakhtenbrot}, B., {The
  correlation of narrow line emission and X-ray luminosity in active galactic
  nuclei}. 2006, {\it \aap}, {\bf 453}, 525, DOI: 10.1051/0004-6361:20054203

\bibitem[{{Netzer} {et~al.}(2004){Netzer}, {Shemmer}, {Maiolino}, {Oliva},
  {Croom}, {Corbett}, \& {di Fabrizio}}]{2004ApJ...614..558N}
{Netzer}, H., {Shemmer}, O., {Maiolino}, R., {et~al.}, {Near-Infrared
  Spectroscopy of High-Redshift Active Galactic Nuclei. II. Disappearing
  Narrow-Line Regions and the Role of Accretion}. 2004, {\it \apj}, {\bf 614},
  558, DOI: 10.1086/423608

\bibitem[{{Raki{\'c}} {et~al.}(2017){Raki{\'c}}, {La Mura}, {Ili{\'c}},
  {Shapovalova}, {Kollatschny}, {Rafanelli}, \&
  {Popovi{\'c}}}]{2017A&A...603A..49R}
{Raki{\'c}}, N., {La Mura}, G., {Ili{\'c}}, D., {et~al.}, {The intrinsic
  Baldwin effect in broad Balmer lines of six long-term monitored AGNs}. 2017,
  {\it \aap}, {\bf 603}, A49, DOI: 10.1051/0004-6361/201630085

\bibitem[{{Shen} {et~al.}(2011){Shen}, {Richards}, {Strauss}, {Hall},
  {Schneider}, {Snedden}, {Bizyaev}, {Brewington}, {Malanushenko},
  {Malanushenko}, {Oravetz}, {Pan}, \& {Simmons}}]{2011ApJS..194...45S}
{Shen}, Y., {Richards}, G.~T., {Strauss}, M.~A., {et~al.}, {A Catalog of Quasar
  Properties from Sloan Digital Sky Survey Data Release 7}. 2011, {\it \apjs},
  {\bf 194}, 45, DOI: 10.1088/0067-0049/194/2/45

\bibitem[{{Shields}(2007)}]{2007ASPC..373..355S}
{Shields}, J.~C., {Emission-Line versus Continuum Correlations in Active
  Galactic Nuclei}. 2007, in Astronomical Society of the Pacific Conference
  Series, Vol. {\bf  373}, {\it The Central Engine of Active Galactic Nuclei},
  ed. L.~C. {Ho} \& J.~W. {Wang}, 355

\bibitem[{{Sulentic} {et~al.}(2000{\natexlab{a}}){Sulentic}, {Marziani}, \&
  {Dultzin-Hacyan}}]{2000ARA&A..38..521S}
{Sulentic}, J.~W., {Marziani}, P., \& {Dultzin-Hacyan}, D., {Phenomenology of
  Broad Emission Lines in Active Galactic Nuclei}. 2000{\natexlab{a}}, {\it
  \araa}, {\bf 38}, 521, DOI: 10.1146/annurev.astro.38.1.521

\bibitem[{{Sulentic} {et~al.}(2000{\natexlab{b}}){Sulentic}, {Zwitter},
  {Marziani}, \& {Dultzin-Hacyan}}]{2000ApJ...536L...5S}
{Sulentic}, J.~W., {Zwitter}, T., {Marziani}, P., \& {Dultzin-Hacyan}, D.,
  {Eigenvector 1: An Optimal Correlation Space for Active Galactic Nuclei}.
  2000{\natexlab{b}}, {\it \apjl}, {\bf 536}, L5, DOI: 10.1086/312717

\bibitem[{{Zhang} {et~al.}(2013){Zhang}, {Wang}, {Gaskell}, \&
  {Dong}}]{2013ApJ...762...51Z}
{Zhang}, K., {Wang}, T.-G., {Gaskell}, C.~M., \& {Dong}, X.-B., {The Baldwin
  Effect in the Narrow Emission Lines of Active Galactic Nuclei}. 2013, {\it
  \apj}, {\bf 762}, 51, DOI: 10.1088/0004-637X/762/1/51

\end{thebibliography}
\end{document}